# Preliminary Study on the RF tuning of CSNS DTL

Xuejun Yin[1,2](殷学军), Ahong Li[1] (李阿红), Yongchuang Xiao[1] (肖永川), Qiang Chen[1] (陈强),
Huachang Liu[1](刘华昌), Keyun Gong[1] (巩克云), Shinan Fu[1](傅世年)

Institute of High Energy Physics, Chinese Academy of Sciences, Beijing 100049, China

Institute of Modern Physics, Chinese Academy of Sciences, Lanzhou 730000, China

**Abstract:** In the R&D of the CSNS Drift Tube Linac (DTL), the first unit tank with 28 drift tubes has been developed. The axial accelerating field is ramped from 2.2MV/m to 3.1MV/m in this tank. The required field flatness is less than ±2 % with the standard deviation of 1 % for the beam dynamics. And the field stability should be less than 1% for machine stable operation. After the successful alignment, the RF tuning was carried out focusing on the field profile measurement. Four slug tuners and 11 post couplers were applied in this procedure. The ramped filed and required stability had been achieved by fine adjustment of the slug tuners and post couplers. In this paper, the preliminary tuning results are presented and discussed.

## 1. Introduction

The CSNS Alvarez DTL working at 324MHz will accelerate the H$^-$ ion from 3.0Mev to 80.0 MeV[1]. The first unit tank (~2.8m in length) was consisted of 29 accelerating cells including 28 full drift tubes and two end plates. In this tank, the axial accelerating field is designed as ramped from 2.2MV/m to 3.1MV/m. To make the DTL tank operate stably and against any perturbation caused by machining errors and the small deformation due to thermal load, slug tuners and post couplers were utilized for the RF tuning procedure. For our case, 4 slug tuners and 11 post-couplers were used.

The field flatness is required ≤±2 % with the standard deviation of 1 % based on the beam dynamics consideration. Noting that the tuning was performed in the air but the machine is operated under the vacuum, so the air effect should be compensated. The estimated frequency change due to the evacuation is about -100 kHz[2]. In addition, difference between the tuning temperature (21 ℃) and the operating temperature ( 26℃ ) can provide an additional frequency change of 20 kHz when the frequency sensitivity on the temperature is assumed to be about 4 kHz/℃[3]. Therefore, the target tuning frequency is set to be about 323.920 MHz ±5 kHz at 21℃. Furthermore, the maximum local frequency shift is expected to be not larger than 8 kHz, which corresponds to the local temperature perturbation of about 2℃. The beam loading can also cause the frequency shift, but will be less than 2 kHz. Therefore, the required stability is determined to be less than 100%/MHz, which means that the field stability sensitivity is less than 1 % with 10 kHz perturbation.

To operate DTL effectively, basic parameters such as the resonant frequency, field flatness and field stability need to be tuned by following the requirements including four steps: 1st, Field flatness tuning with the slug tuners; 2nd, Field stability adjustment carefully with the post couplers; 3rd, Field flatness recovery by rotating the post couplers; 4th, Resonant frequency adjustment by moving all slugs uniformly, where accurate measurements of the field profile are crucial.

## 2. Perturbation measurement

### 2.1 Field profile measurement with bead-pull method

In the DTL tank, the field profile can be measured with the bead-pull method based on the Slater's perturbation theorem [4]. It has been shown that the change in resonant frequency due to the bead

perturbation is proportional to the relative change in stored energy:

$$\frac{\Delta\omega}{\omega_0} = \frac{\Delta U}{U} = \frac{-\int_{V_0}(\Delta\epsilon E \cdot E_0^* + \Delta\mu H \cdot H_0^*)dv}{\int_{V_0}(\epsilon_0 E \cdot E_0^* + \mu_0 H \cdot H_0^*)dv}. \quad (1)$$

where $V_0$ is cavity volume, E and $E_0$ are the perturbated and unperturbated electric field, H and $H_0$ are the perturbated and unperturbated electric field d magnetic field respectively.

For the case of a small sphere bead with radius r, where the perturbed field can be approximated by unperturbed field, it can be shown that:

$$\frac{\Delta\omega}{\omega_0} = -\frac{\pi r^3}{U}\left[\frac{\epsilon_r - 1}{\epsilon_r + 2}\epsilon_0|E_0|^2 + \frac{\mu_r - 1}{\mu_r + 2}\mu_0|H_0|^2\right]. \quad (2)$$

By noting that magnetic field strengths on the z axis are zero since the DTL tank is operating in TM010 mode, then Eq. (2) reduces to

$$\frac{\Delta\omega}{\omega_0} = -\frac{\pi r^3}{U}\left[\frac{\epsilon_r - 1}{\epsilon_r + 2}\epsilon_0|E_0|^2\right]. \quad (3)$$

For a spherical metal bead ($\epsilon_r \to \infty, \mu_r \to 0$), Eq. (3) can be simplified furthermore to

$$\frac{\Delta\omega}{\omega_0} = -\frac{\pi r^3 \epsilon_0}{U}|E_0|^2. \quad (4)$$

Here, the frequency shift due to the spherical bead is proportional to the square of the field intensity on the spot where the bead is located. With this relation, the field profile can be obtained by measuring the frequency shift as the bead is moved through the cavity.

**2.2 DTL tuning setup**

Fig.1 (a) shows the schematic diagram of the field-measuring apparatus of DTL tank 1 based on the Slater's perturbation method. The spherical hollow bead with 3.0 mm diameter is used. The bead transferred by a step motor and the speed is adjusted to complete the measurement in three minutes, which gives the spatial resolution of about 0.5 mm. The phase shift is measured using the vector network analyzer and the computation is carried out on the computer so that the derived field can be obtained, integrated, plotted and compared. The error caused by bead shift from the axis is estimated using the field distribution near the axis. The bead center shift up to ±0.2 mm including the sag when it moves is acceptable.

Fig.1 (b) shows the photo picture of the overall tuning setup with tuning elements such as slug tuners, post couplers and RF pickups.etc. During the flatness tuning and stability tuning, the movable aluminum slugs and the aluminum post couplers are used.

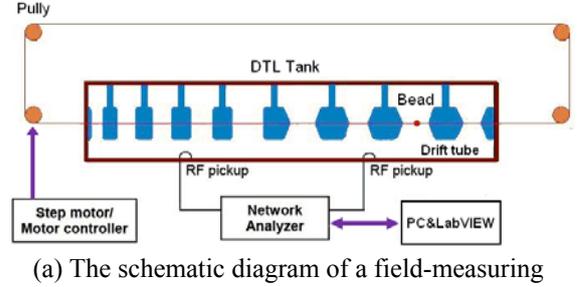

(a) The schematic diagram of a field-measuring

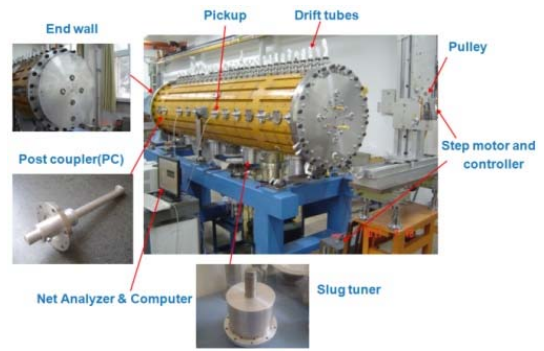

(b) The apparatus photo of DTL RF tuning
Fig.1. DTL RF tuning setup

## 3. RF Tuning results
### 3.1 Field flatness tuning with slug tuners

Equation (4) in the previous analysis shows that the local resonant frequency shifts are proportional to the local electric field energy in opposite direction. And the local resonant frequency can be modified by changing the insertion length of the slug tuner. Therefore, it is feasible to tune the DTL tank to have a ramped field profile by adjusting the position of the movable slug tuner.

As mentioned above, the required field flatness is ±2 % with the standard deviation of 1 %.To tune field flatness effectively, the frequency changes due to the perturbation caused by the slug tuners should be measured. The four slug tuners are initially set in constant length in the tank at 50mm.The measured frequency spectrum is shown in Fig.2. As shown in this figure, the working frequency is 323.4517MHz which corresponds to the TM010 mode. We also measured the two nearest neighbor frequencies which are 319.437MHz and 327.315MHz, and the frequency intervals Δf correspondingly are 4.01MHz and 3.86MHz respectively.

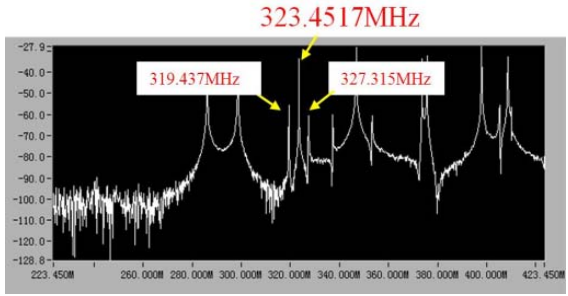

Fig.2.The frequency spectrum

The effect of slug tuner's position on the resonant frequency is measured carefully and plotted in Fig. 3. We can find that the resonant frequency shift is linearly proportional to the insert depth of the slug tuner, especially at the range of 10~80mm. And the measured sensitivities for each slug are ranged from 1.6 to 6.7 kHz/mm. Furthermore the frequency shifts caused by the insert length of tuner are different depend on the positions of the tuners in the tank. The tuner1 at low energy end only has caused 100 kHz shift but the tuner4 at high energy end has 670 kHz.

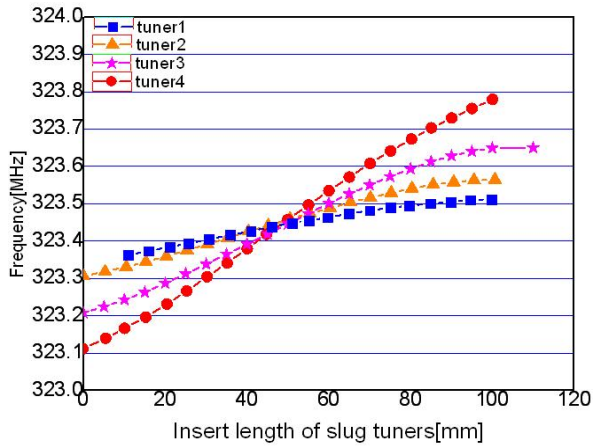

Fig.3.The frequency shift caused by slug tuners

After fine tuning by slug tuners in several iterations, the required field flatness was achieved. The field profiles before and after the slug tuning are plotted in Fig.4. As can be seen, the field flatness decreases from 11.46% to 2.21% after tuning.

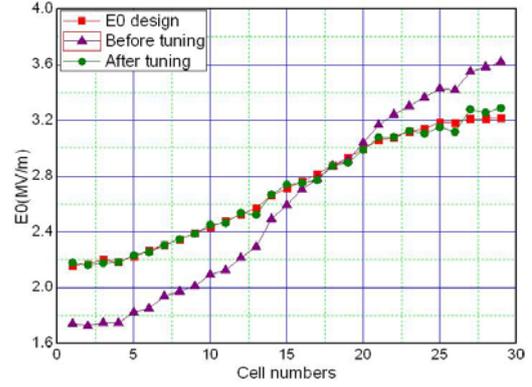

Fig.4.The field profiles before and after tuning

### 3.2 Field stabilization tuning by post couplers

The RF structure with many cells like DTL is very sensitive to the perturbation such as fabricating errors and the small deformation due to power load on the cavity wall, which would cause a large change in the field profile. To make the accelerating field profile insensitive to such perturbation, post couplers (PCs) are used in DTL cavities to create a secondary coupled resonator system, which can form the post coupler modes (PC modes) in the frequency spectrum at a lower frequency band than the operating frequency 324MHz.

Initially, the post couplers are inserted as far as possible. The frequency spectrums of PCs mode are much lower than the operating mode frequency. After then, the post couplers are gradually extracted and the PC frequency spectrum will move from the low band to high band during this extraction. When the PCs highest mode PC1 is adjusted to near the symmetric position in the frequency spectrum to the first higher order mode of TM mode such as TM011 with respect to the operating modeTM010, the perturbation effects from both post coupler modes and higher order modes cancel each other, which will lead to the field stabilization[5].

So, after obtaining the flat field profile, the field stabilization tuning is performed necessarily. To measure the degree of the field stabilization quantitatively, the stability parameter- Tilt Sensitivity (TS) [6] is defined as:

$$TS[\%/MHz] = \frac{E_m - E_0}{E_0} \times \frac{1}{\Delta f[MHz]} 100\%. \quad (5)$$

Where $E_m$ and $E_0$ are perturbed and unperturbed electric fields respectively, and $\Delta f$ is the amount of frequency perturbation. In our measurement, two kinds perturbation such as ±20kHz are introduced by extracting and inserting two movable slug tuners located at the both ends of the tank, as shown in the Fig5. Both of perturbation make about 2% tilt of the field distribution without post couplers.

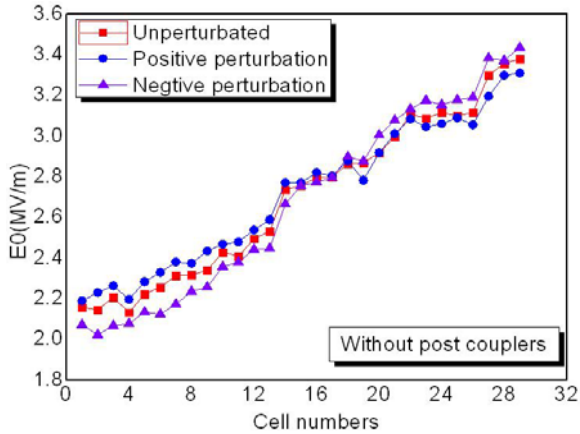

Fig.5.The field profile with and without perturbation

When the field stabilization is performed by tuning with the post couplers, all post couplers are set at the same position initially and then gradually extracted outward until the required tilt sensitivity is obtained. Fig6 was the measured frequency spectrum when the stability was achieved.

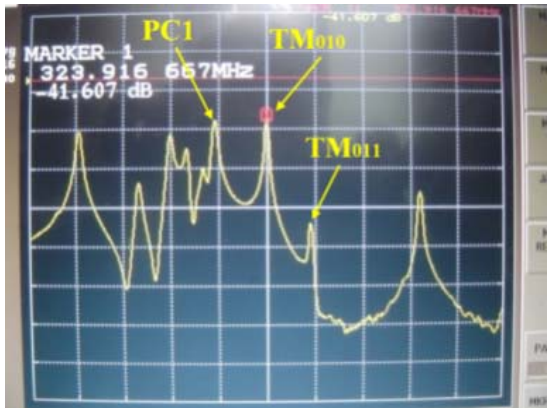

Fig.6. The measured frequency spectrum

As can be seen in Fig. 6, the post coupler modes are located in the lower band with respect to the operating modeTM010. In our tuning results, both of the frequency spacing of PC1 mode and TM011 mode with respect to the operating mode TM010 are ~5.82MHz. The measured dispersion curves of TM modes and PCs modes after getting field stability are shown in the Fig.7. The system, composed by two chains of coupled resonators, has two bands of frequencies: the TM band and the PC band. Since the first unit DTL tank has 11 Post-Couplers, in the PC band there are 11 resonating modes correspondingly.

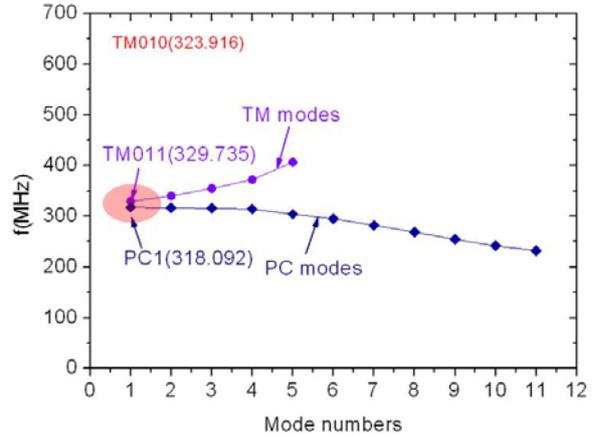

Fig.7.Dispersion curves of TM and PCs modes

It should be pointed that the flatness of the field profile would be somewhat distorted during the field stabilization, which can be recovered by rotating the tip attached at the end of each post couplers. Of course the rotating angles and direction of each tips are not same.After then, the field flatness and the stability were tuned by tuning the slug tuners and post-couplers iteratively. This completes the coarse tuning and the fine tuning is conducted with the tilt sensitivity measurement. Finally, the field flatness profiles with and without perturbations are obtained as presented in Fig.8. The tuning results are summarized in table 1.The field flatness was less than 1.89%(≤2% in design) with standard deviation of 0.87 %, and the stability error decrease from 250~-165%/MHz to 99~-62% /MHz corresponding2.5~-1.65% to 0.99~-0.62%.

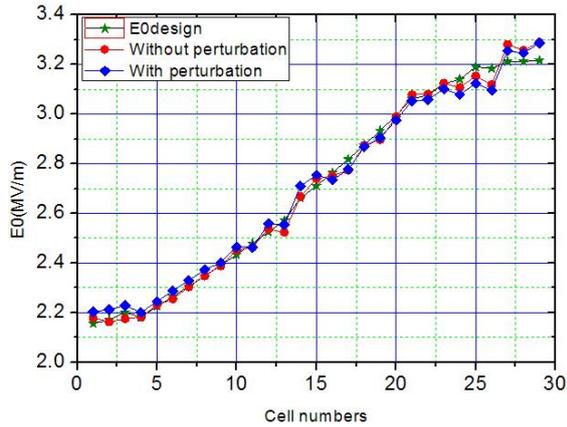

(a) Field flatness

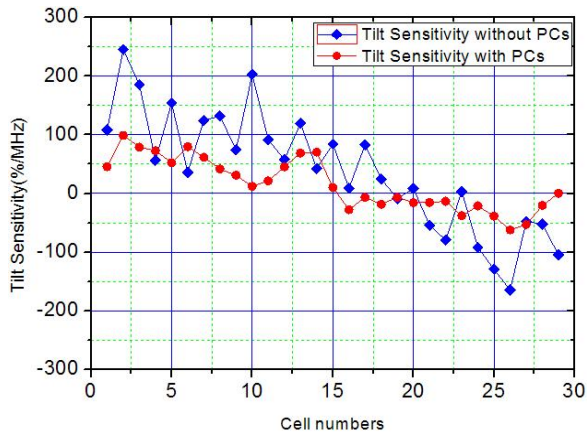

(b) Tilt Sensitivity

Fig.8.CSNS DTL1 RF tuning results for cold measurement

Table 1. The CSNS DTL1 RF tuning results

|  | Designed | Measured |
|---|---|---|
| Frequency /MHz | 323.920 ±5 kHz | 323.917 |
| Field flatness | ≤±2 % | <1.89 % |
| Tilt sensitivity | ≤1 %(100%/MHz) | 0.99~-0.62% |

## 4. Conclusions

In the R&D of CSNS Alvarez DTL, the RF tuning measurement was carried out carefully. The field-measuring apparatus based on the Slater's perturbation theorem was set up. The main measurement procedure includes the frequency measurement, the field flatness tuning and the field stability tuning. The measured sensitivities were from 1.6 to 6.7 kHz/mm for four slug tuners. After some iteration fine tuning, the field flatness ≤1.89% and the tilt sensitivity <1% were successfully achieved. In the near future, the frequency perturbation caused by slugs, RF pickup and RF power couplers should be considered in the stability parameter. And the error of the measurement system should not be neglected. Finally it should be beneficial to improve the environment within the RF tuning measurement system.